\documentstyle[12pt,makeidx,epsfig,wrapfig,amsbsy,amsfonts]{article}
\begin{document}
\topmargin=-1cm
\oddsidemargin=0cm
\textwidth=16cm
\textheight=21.5cm
\raggedbottom
\sloppy

\newcommand{\ee}{\mbox{$e^{+}e^{-}$}}
\newcommand{\er}{\mbox{$\pm$}}

\begin{titlepage}
\pagenumbering{arabic}

\flushright{IHEP 2002-31}
\begin{center}
{\bf The scintillating crystal detector for the  Dark Matter
          searches (Proposal) }
\bigskip
\bigskip
\bigskip
\medskip
\end{center}

\begin{center}
 AKOPDZHANOV G.A., BLICK A.M., KOZELOV A.V., POGOSOV V.S.$^{\ast}$,
 RYKALIN V.I., SHAPKIN M.M, TCHIKILEV O.G.\\

\bigskip
\bigskip
\bigskip
\begin{abstract}
\noindent
The study of the crystal scintillating detector
( based on pure $CsI$ ) is proposed with the aim of use in future
 Dark Matter searches. The planned  energy range of the recoil nuclei
detection between  2 and 40 $keV$ allows to register
neutralino interactions  with masses up to 100 $GeV/c^2$.

\end{abstract}
\bigskip
\bigskip
\bigskip
\bigskip
\medskip
  {\em $\ast$ Yerevan Physical Institute (Armenia) \/}
\end{center}
\vfill
\vfill
\vspace{\fill}
\end{titlepage}

\pagebreak

\setcounter{page}{1}


\begin {center}
{\bf
 Short review of the Dark Matter search experiments
}
\end {center}

 Recent astrophysical studies  indicate that the part of the matter
in the Universe is of  a non-baryonic nature.
 The cosmological models give satisfactory  description of the formation
 process of the spiral galaxies under assumption that nearly $70\%$ of its
 mass is due to the so called ``Cold Dark Matter'' (CDM). It is generally
 assumed that the dark matter halo is present in our Galaxy with
the density around $0.3~GeV/cm^{3}$  in the vicinity of the Solar System.
It is usually assumed that the invisible dark matter forms
a nonrotating spherically symmetric cloud in the center of the Galaxy with
the  Maxwellian distribution of the velocities.

The most probable candidates for such dark matter halo are Weakly
Interacting Massive Particles ($WIMP's$), predicted in
supersymmetric models. The recent searches of these particles in
the frame of the Minimal Supersymmetric Model (MSSM) at the
Tevatron and LEP colliders still have no positive results. An
upper limit for the mass  of the lightest neutralino (a mixture of
photino, zino and higgsino) is $m_{\chi_1^0} > 32.5~GeV$ [1].

In addition to the accelerator experiments one can look for
elastic and inelastic interactions of the relic $WIMP's$ with the
detectors material. Under  the different assumptions the collision
frequency can  vary from 0.001 to 1~$\frac{events}{kg\cdot day}$~[2].

Around twenty neutralino search experiments are now working or under
the planning. There are two different groups of the experiments.
The low mass detectors look for the scatterings
on detector nuclei which are not explained  by the background radiation.
The high mass detectors look for  the seasonal variations of the counting
rates possibly connected with  the Earth circulation around the Sun.

The value of the minimum collision frequency can give us the
estimation of the background conditions for an "ideal" neutralino
registering detector. Let us assume that background and useful
event rates are equal, this gives  the background rate level
around 0.00001~$\frac{events}{kg\cdot day\cdot keV}$ in the
$\sim100~keV$  recoil nuclei energy range. Surely this background
 level rate is unreachable.

The data analysis of the current experiments shows that the main
background contributions are radioactive admixtures in detector
materials and its surrounding and also neutrons in the $MeV$-range
originating from high energy interactions of cosmic muons. The low
background setup needs special conditions --- underground
laboratories, very low radioactivity of the  materials in use,
powerful cosmic radiation shielding, etc. Various physical methods
of background rejection are also used.

  One of the ways to discriminate the background  is to use  the
different ionization losses of electrons and recoil nuclei in the
same dense materials. For instance an electron signal in a
$Ge$--counter exceeds several times  a signal given by recoil
nucleus. As a rule, the phonon signal depends on  the total energy
release of a particle. Such a method of the background suppression
with the $Ge$--detector at $mK$--temperatures was used in the
CDMS~[3] setup, where authors found 3 events treated as CDM
elastic scattering.

It should be be stressed that the  background level achieved for
low mass detectors is of the order 0.01~$\frac{events}{kg\cdot
day\cdot~keV}$~[4] for recoil nuclei energy  $10 - 100$~$keV$.
This background sharply increases if one register particles with
energies less than $5$~$keV$. The better background rejection than
indicated above is planning to achieve in the DRIFT setup~[5]. This
experiment is starting in the nearest time with the aim to look for
$WIMP's$ with masses $\sim 1000$~$GeV$.

Among the current experiments the highest mass of the active
detector medium is used by the DAMA ($\sim100~kg$)~[6,7] and
ELEGANT~V (662~kg)~[8] experiments (they used crystals of
$NaI(Tl)$). Other experiments using the same technique have much
smaller weight of crystals. The main goal of these experiments is
to measure  recoil nuclei energy spectra which can be used for
extracting of the neutralino interaction cross--section and mass.
 The signature of Dark Matter in these experiments is the
seasonal modulation of the signal.

 The total background level of such detectors is usually two order
of magnitude more than compared to cryogenic experiments, but it
can be  significantly lowered using extremely pure
 materials. The special technique is used to reduce
the natural radioactivity of crystals and photomultipliers (PMT).
Each crystal of the DAMA detector ($\sim10~kg$) is viewed by two
photomultipliers from opposite sides (the noise frequency of each
PMT is about 0.1~$kHz$). During the data taking the pulse shape of
each PMT is stored with the following discrimination. The
discrimination is based on the fact that the recoil nucleus have a
much shorter scintillating time than the electrons. The high
temperature stability is required to analyze the pulse shape --- it is
better than 1$^{\circ}C$ in the DAMA detector. In the latest DAMA
publications $2~keV$ energy threshold for X-rays is given (taking
into account quencher-factor the threshold for recoil nuclei is
about $10~keV$). One should mention that they don't present
analysis of the efficiency of the pulse shape discrimination near the
registration threshold. Using seasonal variations analysis they
obtain neutralino mass equal to $40\div 50~GeV$ and cross-section
$5\div 7~pb$ per nucleon~[7].

 Experiment ELEGANT~V has made one year exposure of the $NaI(Tl)$
crystal setup. Some modulation at the level of statistical
fluctuations has been observed~[9]. The main background in this
experiment is due to the high concentration of the $^{222}Rn$ in
the underground laboratory. It is planned to move the setup to the
new underground hall with a lower radon contamination.

 Another high mass detector experiment of the CRESST~[10]
collaboration is planning to carry out  one year exposure of the
"heavy" scintillator $CaWO_{4}$ (total weight 100~kg) at the
liquid helium temperature $12~mK$. Each crystal is viewed by both
a photodetector and a bolometer to obtain a sufficient background
rejection. The background rejection of electrons
 of $\sim99.9 \%$ at the 15~$keV$ is planned. This experiment is
expected to provide the background rate
level about 1~$\frac{events}{kg\cdot day\cdot keV}$.

  The main advantage of using the scintillating crystals ($NaI(Tl)$, 
$CsI(Tl)$ etc.) in dark matter search experiments is relatively low cost 
and simplicity of setup. This gives a possibility to construct an
apparatus with larger working mass. From our point of view there
are two significant drawbacks in the existing experiments (DAMA,
ELEGANT~V). One of them is their inability of full background
rejection in the experiment. Another one is a relatively high energy
registration threshold if compared with $Ge$ detectors. For
example, usage of only two PMTs for the large
($10\times10\times100~cm^{3}$) scintillating crystal in ELEGANT~V
leads to a high threshold because geometric acceptance is
significantly less than $4\pi$. Random coincidences of PMT signals
during luminescence decay time  give contribution into the background
even with a low single electron PMT noise level and their
suppression decreases the efficiency of low-energy recoil nuclei
registration.

It should be mentioned that in cryogenic detectors
(where ionization and thermal signals are compared) the registration
threshold is about several hundreds $eV$, but electrons and recoil
nuclei could not be distinguished at $\sim 3~keV$ and reliable
discrimination in different setups starts only at $10-15~keV$.

 Also the recoil discrimination by the electronic analysis of the pulse shape,
as used in the DAMA experiment, is questionable for small signals.
Indeed, let us assume that the scintillator with PMT is exposed to
the pulsed neutron and gamma ray sources and the pulse shape is
registered (triggering from  the pulsed source). With the signal
level around $10^{3}$ photoelectrons per pulse we have for nuclei
and for electrons two different luminescence decay times of the
scintillator. In another extreme case when the signal level is
about one photoelectron per pulse of the source and measuring
arrival times of the photoelectrons we need near $10^3$ pulses in
order to measure just the same curves. This simple example shows
that for the luminescence decay times of the scintillator and the
real integration times of PMT signals the pulse shape
discrimination works only above some threshold number of
photoelectrons (or above some threshold recoil energy).

\begin{figure}
\epsfig{file=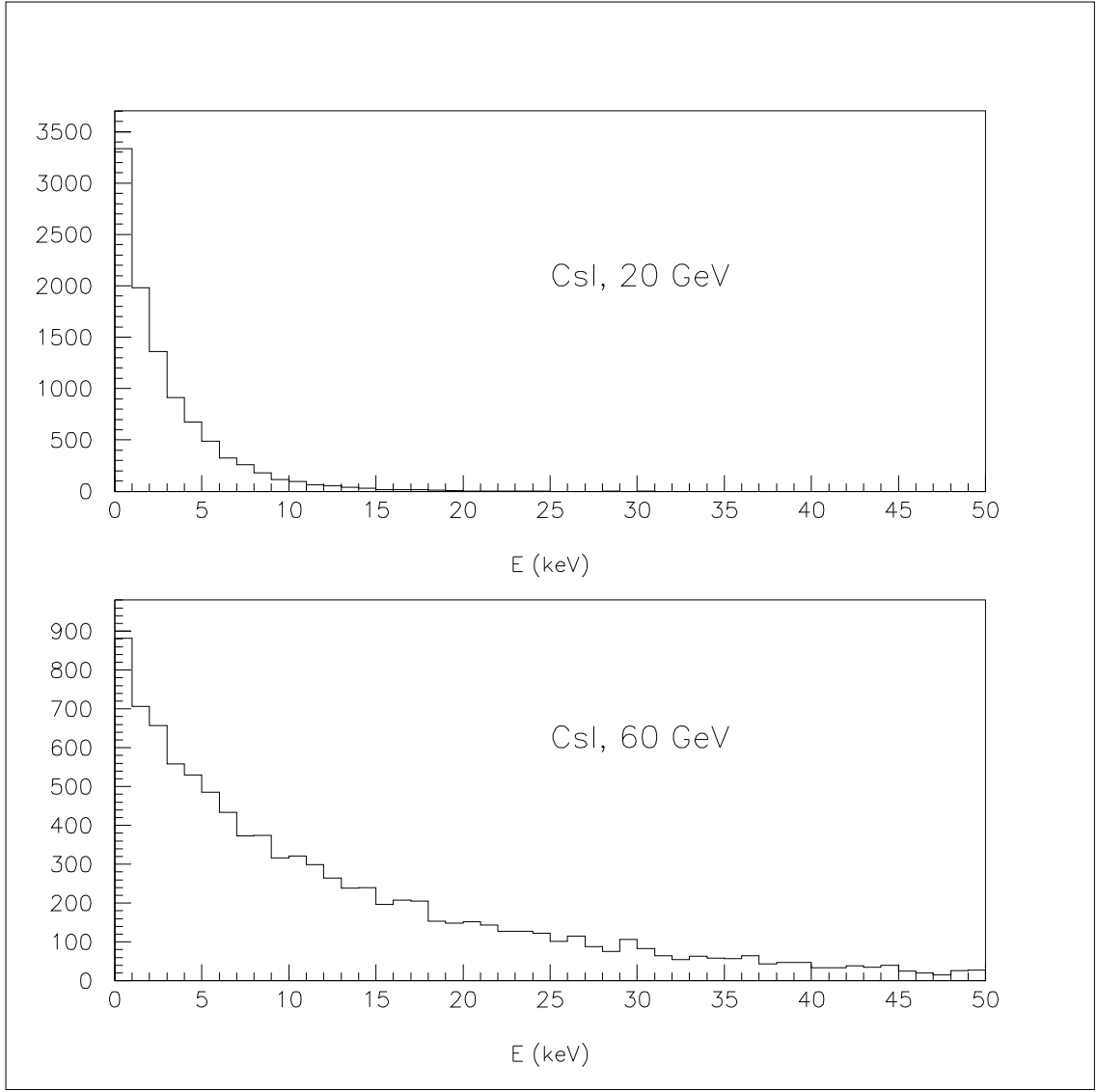,width=15cm}
\caption{ The energy spectrum of recoil nuclei in the neutralino
 elastic scattering on the $CsI$.}
\end{figure}

 So, some progress in the experimental searches of the Dark Matter in the
mass region below 100~$GeV$  can be achieved  with cryogenic
devices (as CRESST for instance), or using increased detecting
mass in DAMA, or a more background rejection in ELEGANT V. If
seasonal variations are not detected, using detector mass and
background level one could set an upper limit for the neutralino
interaction cross-section. The ambitious solution for the
determination of the upper limit of the cross-section is proposed
in project with one tonne of enriched $^{76}Ge$ as working
material~[4]. Another way of solution of this problem is
the construction of a high-precision setup with a cheaper detecting
material.

\begin {center}
{\bf
     Main goals of the project
}
\end {center}

 We propose to build a relatively inexpensive detector with pure $CsI$
 for the relict neutralino detection.

 $CsI$ could be used because of nonhydroabsorbtion, easy mechanics
operations, smaller quenching-factor and the sharp dependence of
light yield on the temperature of a crystal.
 At the temperature of liquid nitrogen its light yield is increased by a
factor of 10 with simultaneous increase of light decay time~[11,12].
Moreover, $CsI$ crystals have better radioactive background
rejection~[13] when compared to $NaI(Tl)$, i.e. are more suitable
for the detection of recoil nuclei at low energies. Because almost
equal atomic weights of $Cs$ and $I$ there is a possibility that
their quenching-factors are close, and Monte-Carlo simulation of
energy spectra of these nuclei could be simplified.

 The energy spectrum of recoil nuclei in the neutralino
elastic scattering on the $CsI$ for several values of $WIMP's$
masses with Maxwellian velocity spectrum is shown in fig.1.
One can observe that quite good efficiency (or number of
photoelectrons per $1~keV$ of the absorbed energy) and recoil
nuclei discrimination must be reached in the recoil energy region
lower $40~keV$.

 During detector design we take into account the following
considerations. For the rough estimation we take  mean energy for one
photoelectron production $\sim 1.5~keV$. In
this case the upper registration limit is about 30 photoelectron.
Therefore, optimal detection mode is measurement of arrival times
of single photoelectron pulses and their subsequent summation. The
upper registration threshold will approximately correspond to
energy deposition of a single X-ray quantum with energy of $5.9~keV$
from $^{55}Fe$ source. Signals larger than this threshold are
treated as the background and rejected during data processing.

 We suppose to reach good efficiency using optimal choice of
crystal size $(\sim 10 \times 10 \times 10 ~cm^3)$ and geometric
acceptance about $4\pi$.  Therefore we choose the following design
of the modular detector. Planes of a scintillating crystal are
observed by six PMTs via quartz light guides. Crystal and PMT
input windows are situated inside a copper cassette cooled to the
liquid nitrogen temperature. The cassette is put into the center
of a shielding box, the function of which is to protect from
the surrounding natural radiation.

 The employment of six PMTs per crystal increases registration
efficiency and simplifies time analysis of single photoelectron
signals. The detector cooling reduces PMT thermoemission noise and
the frequency of random coincidences is decreased in spite of
increased number of PMTs. The PMT noise due to extract of group of
photoelectrons from the photocathode could be rejected by the
amplitude analysis.

We estimated the production energy of a single photoelectron
at the temperature of liquid nitrogen
as $500~eV$ for recoil nucleus (taking into account
the rise of quencher-factor with the decrease of recoil nucleus
energy~[13]). Four detected photoelectrons correspond to recoil
nucleus energy about $2~keV$. This is the lower limit of the 
measured recoil nucleus energy.

  The $40~kev$ upper limit means that the electrons and
$\gamma$-quanta from the external radiation with energy deposition in
crystal less than $6~keV$ will simulate neutralino interactions
and give the main  contribution into the background. Another
background source is interactions of neutrons from the surrounding
with the scintillator.

  We propose to take into account the background from the external
radiation using a background detector. To make a such detector we
simply replace the $CsI$ crystal by a plastic scintillator. In
this case we can separate to a certain degree interactions of
neutrons and $\gamma$-quanta. A neutron slowing down in the
scintillator could produce two light pulses which correspond to
interactions with plastic. The time between pulses is in
accordance with the free flight time at a given energy. As plastic light
decay time at the liquid nitrogen temperature decreases to several
nanoseconds, a possibility of separation will appear. The shortening
of the light decay time and the decrease of the PMT noise reduce the
probability of accidental coincidence. In this case the total time
spectrum of six PMTs of the background detector is the sum of the PMT
noise spectra and signals caused by interactions of low-energy
electrons and neutrons with the plastic scintillator.

It would be ideal to achieve the equal light yields from the $CsI$
crystals and plastic scintillator for electrons with energies lower
than 6 $keV$ by variation of the detector
temperature. But even if this goal could not be achieved, 
the measurements of energy losses of
$\gamma$-quanta in the organic scintillator will give  a possibility to
reconstruct energy spectrum of the external $\gamma$ background. To
achieve this goal characteristics of both modules should be almost
identical.

The time spectra of all PMTs with tight photocathodes are measured
before and after the measurements.

  In energy region of neutralino registration there is also a
background originating form nuclear transitions in scintillating
crystal leading to the production of radioactive isotopes, X-rays
and Auger electrons. Careful measurements of crystals in this
energy range are required to identify the isotopes. However, the
knowledge of some background conditions is an advantage of
the proposed detector when compared to the existing setups
for dark matter searches.

The counting rate of each PMT (less than $1000~Hz$) defines requirements
for electronics. A common timer with time resolution $1~ns$ is used.
Fast amplifiers are installed on the PMT base to amplify anode and
last dynode signals. 
 The anode signals are digitized flash ADC.
The dynode signal is used to form a logic
signal and their arrival times are digitized also.

It should be mentioned that after measurements of real time spectra
the electronics could be optimized.

 There is a plan to research the temperature dependence of light yields 
due to recoil nuclei, electrons and neutrons in the $CsI$ and organic
scintillator.
The calibration of the detecting crystal could be performed with the
$\gamma$-quanta from
$^{55}Fe$ source (energies are $5.9~keV$ and $2.5~keV$). For $CsI$
in the single photoelectron mode one should observe two peaks in energy
spectrum.

 The disadvantages of the proposed detector are a relatively small
scintillator mass and six PMTs per module. The PMTs with small content
of radioactive materials are required, and their cost will define
the cost of the whole detector. It could also turned out that natural
radioactivity of PMT's is one of the main sources of the
background. As for the scintillating crystals, available crystals have
various shapes and cutting them to the cubic shape will lead to
the loss of scintillating material. Besides it's desirable to have
a crystal of the largest possible weight.

\begin{figure}
\epsfig{file=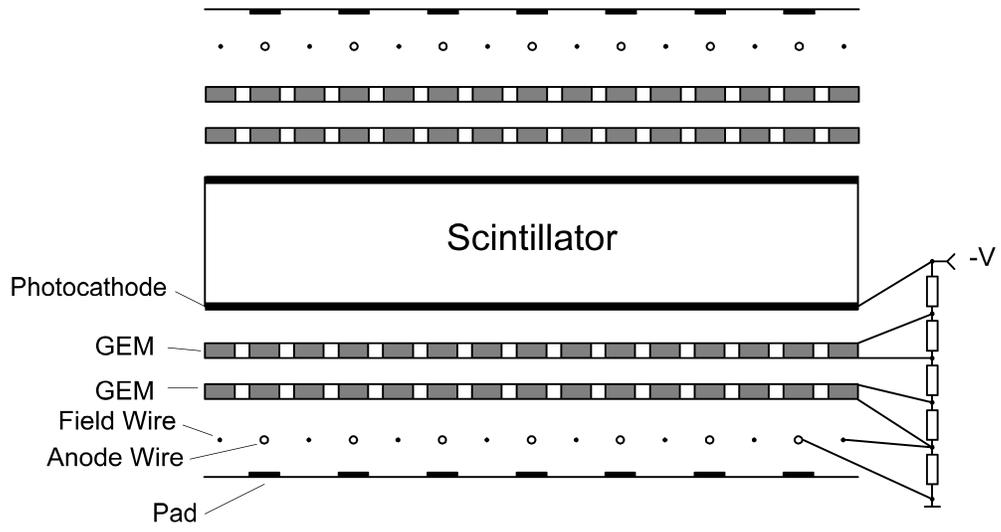,width=15cm}
\caption{ Schematics of the scintillating crystal with GEM
 photomultiplier.}
\end{figure}

For large crystals it will be very promising to use gas electronic
multipliers (GEM) instead of PMTs. Classic schematics of GEM by
F.Sauli is presented in Fig.2~[14].

 Let us consider $10\times 10\times 100~cm^3$ scintillating
crystal. Usually crystals of such a shape are viewed by two PMTs on
small edges. We propose to glue thin quartz plates with visible
light photocathodes to two opposite large edges. After the gas gap two
GEMs with total amplification about 100  are situated. The amplifier
consists of a capton film with small holes in which gas
amplification takes place. Film thickness is $50~\mu m$ and its
production is organized at CERN. The main
amplification takes place near anode wires. Signals from wires and
pads are registered. Gas mixture works in the limited proportional
mode (when there is no transition between the limited mode and 
Self Quenching Streamer mode).
 In this case we will have anode signals with the amplitude
about $30~mV$ and length $\sim 30~ns$ at $50 \Omega$ load [15].

  For the background detector with PMTs the pile-up of signals from
neutrons with small free flight times is very probable. A better
spatial resolution of GEM could eliminate this problem.

  Several groups carry out researches of visible light
photocathodes which will be stable at normal atmosphere
conditions. The best result for quantum sensitivity is $\sim
0.03$~[16]. It's about ten times worse than for PMTs but this loss
will be compensated by better light collection efficiency.

  Replacement of PMTs to GEMs in dark matter detector has obvious
advantages - miserable quantity of passive material, good spatial
resolution and decrease in detector cost.

\begin {center}
     Conclusions.
\end {center}

  We propose the scintillating crystal detector for the Dark Matter
searches. Some advantages of the proposed detector are summarized
below.

1. For $CsI$ the lower energy limit of the recoil nuclei detection will be
$\sim2~keV$ due to the use of the almost $4\pi$--geometry, the liquid nitrogen
working temperature, the single photoelectron regime of the photomultipliers.
This allows effective neutralino interactions search
in the mass range  20--100 $GeV/c^2$.

2. The proposed method of the detector testing allows to carry out
the careful studies of the background conditions and to
estimate an upper limit of the $WIMP's$ interaction cross-section.

3. The modular detector structure allows to increase scintillator mass
in the future.

4. The interaction point of the recoil nucleis (or of the background 
electrons) in the scintillator and
the  dependence of light emission time on particle type will be
taken into account during the data processing.

5. The  electronics of the proposed detector is much simpler
and cheaper when compared  with all existing or planned devices
with analogous aims.

6. For large scintillating crystals PMTs could be replaced by
GEMs. This allows to improve background conditions and reduce
detector cost without loss of efficiency.

\end{document}